# The ghost of AI governance past, present and future: AI governance in the European Union

**Charlotte Stix**[1]

Eindhoven University of Technology, The Netherlands.

…………………

**Abstract:** The received wisdom is that artificial intelligence (AI) is a competition between the US and China. In this chapter, the author will examine how the European Union (EU) fits into that mix and what it can offer as a 'third way' to govern AI. The chapter presents this by exploring the past, present and future of AI governance in the EU. Section 1 serves to explore and evidence the EU's coherent and comprehensive approach to AI governance. In short, the EU ensures and encourages ethical, trustworthy and reliable technological development. This will cover a range of key documents and policy tools that lead to the most crucial effort of the EU to date: to regulate AI. Section 2 maps the EU's drive towards digital sovereignty through the lens of regulation and infrastructure. This covers topics such as the trustworthiness of AI systems, cloud, compute and foreign direct investment.
In Section 3, the chapter concludes by offering several considerations to achieve good AI governance in the EU.

---

[1] PhD Candidate, Philosophy and Ethics Group, Department of Industrial Engineering and Innovation Sciences, Eindhoven University of Technology, The Netherlands. | PhD Student Fellow, Leverhulme Centre for the Future of Intelligence, University of Cambridge. United Kingdom. | Correspondence should be addressed to c.stix@tue.nl.| ORCID iD: 0000-0001-5562-9234.

# Introduction

This chapter will provide a simplified overview of the past, present and future of AI governance in the EU. It will give the reader a solid background understanding of how the EU reached the current status as global leader in the regulation of AI, how all the different pieces are interconnected and where the EU might go next. Section I will discuss a select number of EU policy efforts of the past years and illustrate how they built on each other. It is argued that, by virtue of spearheading 'trustworthy AI', the EU has occupied a position where it has been able to shape the discourse on global AI governance discourse early on. Section II will introduce the history of the EU's AI regulation, proposed in April 2021, highlighting its connection to previous policy efforts, and its roots in adjacent measures to strengthen the EU's technological ecosystem. Finally, in Section III, the author will turn to the future, considering and exploring a number of AI governance areas that are prime candidates to become crucial for AI governance in the EU in the coming decade.

# I. The Past

## Taking stock: the roads towards the EU's AI governance

This section will highlight the most relevant and recent EU policy developments with regard to AI, illustrate how they contributed to shaping both the broader narrative for the EU and how they set the cornerstones for AI governance in the EU. It will be suggested that the AI Act (European Commission, 2021c), the European Commission proposal for a horizontal AI regulation and the accompanying policy measures in the EU form a coherent and strategically aligned link in a chain of policies which were initiated many years ago. To that end, some of these policy documents will be revisited in Section II, which presents the different elements that form the bigger picture of current AI governance in the EU.

While going through the formative policy developments, it is worth underlining that the European Commission has invested in and funded AI and AI-related research and innovation projects for much longer than they have been focussing on the governance of AI, notably, under Horizon 2020 and before. With that in mind, the following paragraphs will set out the main elements that led to the EU to push for ethical governance of AI and to make it their guiding principle for accompanying policy measures.

## I.i. The roads that led us here

The earliest key policy document is the resolution by the European Parliament with 'Recommendations to the Commission on Civil Law Rules on Robotics' in 2017 (henceforth: 'Civil Law Rules on Robotics', European Parliament, 2017). Although not yet referring to AI directly in the title, the resolution laid one of the first cornerstones for the succeeding process from the European Parliament's side, by suggesting that the EU's legal framework should be updated and complemented by ethical principles on the topic, that the environmental impact of AI and robotics should be kept low, and that the societal and economic impacts of future systems deserve heightened attention.[2]

Shortly thereafter, the European Economic and Social Committee (EESC) presented their 'Opinion on AI' (European Economic and Social Committee, 2017). The 'Opinion on AI' (European Economic and Social Committee, 2017) discusses the need to verify, validate and monitor AI, as well as AI-based systems, advocates for an overarching "human-in-command" approach and leans into the necessity for ethical, societal and safety considerations. Accordingly, recommendations cover the development of a code for ethics, a ban on Lethal Autonomous Weapons Systems and the development of suitable standardisation systems for AI.

On evaluation, we can already see that while they are among the earliest EU policy documents on the topic, the EP's 'Civil Law Rules on Robotics' (European Parliament, 2017), and the EESC's 'Opinion on AI' (European Economic and Social Committee, 2017) have some overlaps. These include a demand for inclusion of the ethical dimension in the discussion and an acknowledgement of the societal impact, alongside proposals for recourse.

---

[2] We will revisit the role this Report continues to play in Section III.

The shift towards EU AI governance as a topic largely independent of robotics from the perspective of EU institutions, was further solidified through the European Group on Ethics in Science and New Technologies (EGE) Statement on 'Artificial Intelligence, Robotics and Autonomous Systems' (henceforth: Statement; European Group on Ethics in Science and New Technologies, 2018). An independent advisory body to the European Commission, the EGE advises it on the intersection of science and emerging technologies with ethical, societal and fundamental rights issues. In their 'Statement' (European Group on Ethics in Science and New Technologies, 2018), they echoed the need to establish an overarching framework on AI in the EU with an ethical dimension. The goal would be to tackle the ethical, legal and societal governance issues, ensuring that AI is created with "humans in mind" (European Group on Ethics in Science and New Technologies, 2018). Therefore, in a sense building on the resolution on 'Civil Law Rules on Robotics' (European Parliament, 2017) and 'Opinion on AI' (European Economic and Social Committee, 2017), they proposed the development of several ethical Principles for AI based on fundamental European values. This proposal was both quintessentially European, outlining the importance of fundamental rights and values, and well timed to fit within the broader international landscape, where principles for AI were starting to see their advent (Fjeld et al., 2020; Hagendorff, 2019; Schiff et al., 2020; Zeng et al., 2018).

These three documents together could be seen as the first heralds of where the EU is now: regulating AI with a focus on human-centricity and ethics. They demonstrated what aspects of AI were considered important by the EU's legislative body, the EU's civil society organisation body and the main EU group on ethics at that point. This is a convergence point where demonstrable attention has begun from a legislative angle, a societal angle and an ethical angle. As we will see in Section II this interplay has since continued, and even been strengthened. At this point in time it became strikingly evident that the EU policy space was alert to the challenges posed by AI, as much as to the opportunities AI could hold, and the necessity for a more methodological approach became pressing.

What followed were major leaps. First, the European Commission presented the 'Digital Day Declaration on Cooperation on AI' (European Commission, 2018c) in April 2018. Second, they responded to the call from the European Council to "put forward a human-centric approach to AI"[3] by presenting their AI strategy in the Communication entitled 'Artificial Intelligence for Europe' (European Commission, 2018a), in the same month. I propose that these documents foreshadow current EU AI governance mechanisms. The 'Digital Day Declaration on Cooperation on AI' (henceforth: 'Declaration'; European Commission, 2018c) anticipated the 'Coordinated Plan on AI' (European Commission, 2018b), whereas the 'Communication on AI for Europe' (European Commission, 2018a), in some sense preceded the proposal for a regulatory framework, the AI Act (European Commission, 2021c).

So, how did these documents set out the European Commission approach to AI governance?

In the 'Declaration' (European Commission, 2018c), signed by all 28 Member States in 2018 (at that time including the United Kingdom) as well as Norway, the countries agreed to engage in close dialogue with the European Commission on the topic of AI and coordinate their actions. I propose that this is the first instance internationally where a significant number of countries agreed to coordinate on AI governance.[4] International, later-stage efforts to coordinate among multiple countries such as the Global Partnership on AI (GPAI) or the OECD included the EU by way of representation through the European Commission, as well as a subset of member states. As of this writing, there is no other international AI governance effort that envisages the same level of coordination, alignment of approach, and pooling of resources as the one started with the 'Declaration'[5] and signed off by the member states.[6]

We have seen that the 'Opinion on AI' (European Economic and Social Committee, 2017) called for a "human in command" approach and that the EGE 'Statement' (European Group on Ethics in Science and New Technologies, 2018) called for AI to be made with "humans in mind". This notion of human-centric AI is revisited in the 'Declaration' (European Commission, 2018c), committing signatories to ensure that 'humans remain at the centre of

[3] See: https://www.consilium.europa.eu/media/21620/19-euco-final-conclusions-en.pdf.
[4] Of course, such coordination may be seen as implicit by virtue of these countries being Member States of the European Union.
[5] This Declaration has been expanded on significantly in the Coordinated Plan on AI and its follow up.
[6] It should be noted that the Declaration is non-binding. However, any other international efforts are equally non-binding at the time of this writing.

AI development', and to prevent the "harmful creation and use of AI applications".

Echoing topics outlined in the resolution on 'Civil Law Rules of Robotics' (European Parliament, 2017) and the 'Opinion on AI' (European Economic and Social Committee, 2017), the 'Declaration' (European Commission, 2018c) focuses on the development of a collaborative framework to coordinate on relevant areas such as sustainability, labour market, funding and ethics. Alongside the necessary mitigation of ethical risks, it also touches upon legal and socio-economic risks of AI. All of this is underlined with the recognition that the existing ecosystem needs to be boosted in order for the EU to remain competitive and agile for future challenges.

The 'Communication on AI for Europe' (henceforward: AI Strategy; European Commission, 2018a) picked up on this and presented the EU's three-pronged strategy for AI taking into account the ecosystem, civil society as well as ethics and regulation. In short, it recommended to "(1) boost Europe's technological and industrial capacity; (2) to prepare Europe for the socio-economic changes associated with AI; and (3) to ensure that Europe has an appropriate ethical and legal framework to deal with AI development and deployment" (European Commission, 2018a).

The AI strategy (European Commission, 2018a) also outlined many ambitions that are currently relevant, such as the development of regulatory sandboxes (eventually key for a horizontal EU AI regulation) or a commitment to support centres for data sharing (for example, the EU Data Hubs).

In order to tackle the third pillar of its AI strategy (European Commission, 2018a), the European Commission set up an independent High-Level Expert Group on Artificial Intelligence (AI HLEG) tasked with, amongst other deliverables, the development of Ethics Guidelines.[7]

[7] These will be further explored in Subsection I.ii. Another deliverable not discussed in this chapter are the Policy Recommendations for Trustworthy AI.

Moreover, the AI strategy (European Commission, 2018a) also served to present, for the first time, a clear 'European way for AI' vis-a-vis the international stage. It clearly outlined the role that the EU envisages for itself when it comes to AI governance. It stated that;

> "the EU must therefore ensure that AI is developed and applied in an appropriate framework which promotes innovation and respects Union's values and fundamental rights as well as ethical principles such as accountability and transparency. The EU is also well placed to lead this debate on the global stage. This is how the EU can make a difference - and be the champion of an approach to AI that benefits people and society as a whole."

The EU clearly positioned itself as an actor on the international stage who will put ethical considerations and fundamental rights at the core of AI governance.

Finally, we end this section by pulling some of the threads together while leaving space to investigate the regulatory efforts in Section II. Many of the efforts highlighted culminate or find resonance in the European Commission's 'Coordinated Plan on the Development and Use of Artificial Intelligence Made in Europe' (henceforth: Coordinated Plan; European Commission, 2018b) published in late 2018.

The 'Coordinated Plan' (European Commission, 2018b) picks up where the 'Declaration' (European Commission, 2018c) left off. It too was agreed on by all Member States as well as Norway and Switzerland and is to be updated on a rolling basis. It echoes plans mapped out in the AI strategy (European Commission, 2018a), namely that a European approach to AI should be built upon ethical and societal values derived from the Charter of Fundamental Rights. Moreover, going above and beyond the perspective of previous policy documents, it puts a strong emphasis on what should become the European northstar for AI, by highlighting what it perceives to be interconnected concepts of "trusted AI" and "human-centric AI" (European Commission, 2018b).

The 'Coordinated Plan' (European Commission, 2018b) paints a picture of how the Member States can coordinate their AI strategies, define a common vision and encourage synergies between ongoing efforts in the Member States – with an eye to increasing the EU's global competitiveness and to counteracting fragmentation and competition between like-minded actors. The preliminary framework for coordination homes in on a couple of focus areas such as on commonly shared societal challenges, increased diffusion of AI, support to AI excellence, data availability and a regulatory framework. The last aspect is described as a 'seamless regulatory environment' in other parts of the text, and we will see in Section II what the EU has developed on that front. The 'Coordinated Plan on AI' (European Commission, 2018b) also contained a commitment on the EU's side to invest EUR 20bn into AI by 2020, and scale up towards yearly investments of that sum from then until 2027.[8]

In an adjacent stream, in early 2019, the European Parliament's Committee on Industry, Research and Energy (ITRE) had their report on 'A comprehensive European industrial policy on artificial intelligence and robotics' (European Parliament, 2019) adopted, which articulated a clear need for a "robust legal and ethical framework for AI", and which amongst other things called for ethical principles that are in compliance with relevant EU and national law. Along these lines it also welcomes the work of the AI HLEG, further outlined in Section I.ii. Furthermore, it proposes aspects related to e.g. personal data and privacy, consumer protection, transparency, explainability and bias. In tandem with the EU's approach (as mapped so far), the industrial policy stresses the importance of human-centric technology and to encourage ethical values with regards to AI development and deployment. Indeed, this may set the EU apart and propel it to take lead on an international stage.

Combining the various policy efforts and taking a bird's-eye view, clear directions are emerging that concern the role the EU has set for itself generally and on the international stage. Ethical concerns, fundamental rights and values play a vital role in the EU's AI governance future. To account for this, the next Subsection focuses on the development of ethical principles for AI in the EU – and their subsequent impact.

[8] This includes investment on Member State-level, as well as public-private partnerships, the Digital Europe Programme and Horizon Europe funding (the latter two both run between 2021-2027).

The topic covered in the next Subsection should be seen as an adjacent stream of work that resulted out of the landscape built by the policy initiatives outlined here which became a policy effort in its own right, eventually feeding back into the current landscape mapped in Section II.

## I.ii. Coining 'trustworthy AI'

This Section explores how the EU came to adopt and pioneer the term "trustworthy AI" from its ethical investigations, and what this shift marked.

Subsequent to its ambition to develop an appropriate ethical and legal framework for AI, the European Commission set out to establish two groups to support the AI strategy described in its Communication on AI: the High-Level Expert Group on AI (AI HLEG) and the European AI Alliance. The latter was set up as an accessible online multi-stakeholder platform with the goal of contributing to the work of the European Commission and the AI HLEG.

The AI HLEG, on the other hand, was set up as an independent expert group by the European Commission, populated through a selection process (Stix, 2021). The AI HLEG was tasked with the primary goal of developing ethics guidelines. The result of their work, especially the 'Ethics Guidelines for Trustworthy AI' (AI HLEG, 2019) and the 'Assessment List for Trustworthy AI' (henceforth: Assessment List; AI HLEG, 2020), were core to the recent model of AI governance in the EU as the following paragraphs will demonstrate. To that end, the focal point for the following paragraphs will be on the 'Ethics Guidelines for Trustworthy AI' (A HLEG, 2019) and the associated 'Assessment List for Trustworthy AI: for Self-Assessment' (AI HLEG, 2020), and how the conceptualisation of ethical AI contained in them contributed to Europe's vision of 'trustworthy AI'.

Tasked with the development of ethics guidelines, the AI HLEG, composed of 52 experts representing various sectors and types of expertise, underwent a comprehensive process to take a unique step towards a framework for the ethical governance of AI. Although the work was conducted internally, the AI HLEG did share their progress in meetings open to institutional observers and solicited feedback on their first draft version of the Ethics Guidelines half a year into the process via the AI Alliance[9]. Following the implementation of this public feedback, the AI HLEG presented their final 'Ethics Guidelines for Trustworthy AI' (henceforth: Ethics Guidelines; AI HLEG, 2019) in April 2019.

The 'Ethics Guidelines' (AI HLEG, 2019) constituted the first document that proposed a clear conceptual understanding and framing of what type of AI should be encouraged within the EU. While the document is strongly anchored in EU values and fundamental rights as enshrined in the Charter of Fundamental Rights of the European Union, the core concept is that of 'trustworthy AI'. In this reading, 'Trustworthy AI' is to fulfil three conditions: (i) it should be lawful, (ii) it should be ethical and (iii) it should be robust (both from a technical and social perspective).

While the 'lawful' aspect is left to existing regulation and future regulatory efforts, the document proceeds to outline the other components. In particular, our focus will be on the ethical component. The AI HLEG distilled a number of core values, which informed four principles. These are; *Respect for Human Autonomy*, *Prevention of Harm*, *Fairness* and *Explicability*. From these four ethical principles, they derived their seven key requirements to achieve 'trustworthy AI' and to operationalise these four identified principles.

The seven key requirements covered:

- *Human Agency and Oversight,* which relates to the principle of *Human Autonomy* and requires that AI system's allow for human oversight, support the user's agency and foster fundamental rights.
- *Technical Robustness and Safety,* which relates to the principle of *Prevention of Harm*. It addresses concerns such as resilience to attack (e.g. through data poisoning or model leakage), the need for suitable fallback plans, reliability and reproducibility.

[9] A feedback mechanism was also used in the case of the Assessment List, for which feedback was received through two online questionnaires and through in-depth interviews with different types of organisations.

- *Privacy and Data Governance,* which links to the principle of *Prevention of Harm*. It tackles the initial stages of data collection (e.g. regarding the quality and integrity of the data), as much as the need for data protocols to govern data access, and overall privacy measures throughout the AI life cycle.
- *Transparency,* which links to the principle of *Explicability*. This means, among other things, that traceability should be ensured and that capabilities and intentions (both from a technical POV and from an industry perspective) should be clearly communicated.
- *Diversity, Non-Discrimination and Fairness,* which links to the principle of *Fairness*. It states that all affected stakeholders throughout the AI life cycle need to be taken into consideration and duly involved. This means e.g. ensuring equal access and equal treatment.
- *Societal and Environmental Well-Being,* which relates to both the principle of *Fairness* and the principle of *Prevention of Harm*. It relates to the broadest range of stakeholders, the environment and the wider society. Considerations are e.g. AI's social impact and the sustainability of the current AI supply chain.
- *Accountability,* which is the last key requirement and ties all the previous requirements together. It is informed by the principle of *Fairness*. It focusses on redress mechanisms, trade-offs between principles and the need to have adequate mechanisms in place to report potential negative impacts.

At this point, the threads started with the report on 'Civil Law Rules on Robotics' (European Parliament, 2017), the 'Opinion on AI' (European Economic and Social Committee, 2017) and the AI Strategy (European Commission, 2018a) have come to reach a fuller picture: the EU's ambition to create an ethical approach towards the development and deployment of AI and an appropriate ethical framework has become a reality.

In order to operationalise these key requirements further, the 'Ethics Guidelines' (AI HLEG, 2019) also contained a draft Assessment List which was piloted and revised in the second year of the group's mandate.[10] The final 'Assessment List for Trustworthy AI: for Self-Assessment' (henceforth: 'Assessment List'; AI HLEG, 2020) is the first tool in the EU that took 'trustworthy AI' into account throughout an AI system's lifecycle and outlined how an assessment of that could take shape. It was also among the earliest serious attempts to translate ethical principles for AI into actionable measures for all stakeholders involved throughout the AI lifecycle, be that researchers, industry, government or civil society.

The concept of 'trustworthy AI', in the way that the AI HLEG defined it, became a cornerstone for AI governance in the EU. Building on its previous emphasis on 'human-centric AI' (as we have seen in the European Council's call to the European Commission,[11] the 'Declaration' (European Commission, 2018c) and the 'Coordinated Plan' (European Commission, 2018b)) the European Commission adopted this conceptual approach and expanded upon it in the Communication on 'Building Trust in Human-Centric AI' (European Commission, 2019). In that Communication, the European Commission supports the key requirements and the concept of trustworthy AI, stating that;

> "Only if AI is developed and used in a way that respects widely-shared ethical values, it can [sic] be considered trustworthy."

This can be understood – and as we will see, is reflected in subsequent governance documents and decisions – as the European Commission embracing the concept of 'trustworthy AI' as a core component of its strategic vision.[12]

Equally, it doubles down on the reputation of the European Union as a region that produces "safe and high-quality products" (European Commission, 2019). To consolidate its place as a leader on 'trustworthy AI' on an international stage, the Communication on 'Building Trust in Human-Centric AI' (European Commission, 2019) furthermore launched a consensus building an 'International Alliance for a human-centric approach to AI'.[13] Its goal is to share the EU's vision and ambitions with like-minded international partners.

[10] The European Commission opened a broad stakeholder consultation process where feedback was solicited through three different streams: (a) a quantitative stream, (b) a qualitative stream and (c) a holistic stream. The quantitative stream consisted of two surveys, one for developers and deployers, and one for other stakeholders. The qualitative stream allowed for 50 in-depth day long interviews with selected companies trialling the Assessment list on use-cases. Finally, the last channel allowed for feedback from the broader community, discussion papers, white papers, blog posts and reports were provided alike from entities as broad as individual researchers to international industry.
[11] See: https://www.consilium.europa.eu/media/21620/19-euco-final-conclusions-en.pdf.
[12] It should be noted that the key requirements and Ethics Guidelines are of a non-binding format.

Beyond that, the Communication on 'Building Trust in Human-Centric AI' (European Commission, 2019) further expands on elements of documents such as the 'Coordinated Plan' (European Commission, 2018b). It reiterates the core foci to boost the ecosystem, such as an increase in joint ventures, pooling of data and other building blocks for AI, as well as the strengthening of synergies across Member States. It proposed to launch a set of networks of AI research excellence centres under the Horizon 2020 research and innovation framework programme, to set up networks of AI-focussed Digital Innovation Hubs (DHIs) and to develop and implement a model for data sharing and common data spaces amongst Member States and other stakeholders. These suggestions directly shape the current state of affairs as we will see in Section II.

# II. The Present

## The third way: the EU's AI Northstar

We have now reviewed the recent historical backdrop of the current EU's regulatory strategy, discussing both its roots and predecessors. This brings us to today. The EU is looking to develop an attractive alternative to US and Chinese approaches to AI governance. In order to gain a bird's-eye view of that third way, this Section will highlight a select number of important and current developments, illustrating how they contribute to the EU's direction.

On 19th February 2020, the European Commission published a comprehensive package consisting of: the 'European Strategy for Data' (European Commission, 2020a), the report on 'Safety Liability and Implications of AI, the Internet of Things and Robotics' (European

[13] See: https://digital-strategy.ec.europa.eu/en/funding/international-alliance-human-centric-approach-artificial-intelligence.

Commission, 2020f), and the 'White Paper on Artificial Intelligence: A European Approach to Excellence and Trust' (henceforth: White Paper on AI; European Commission, 2020d). Due to the limited scope of this chapter, we will focus on the 'White Paper on AI' (European Commission, 2020d).

This 'White Paper on AI' (European Commission, 2020d) followed European Commission president von der Leyen's promise in her political agenda to put forward "legislation for a coordinated European approach on the human and ethical implications of Artificial Intelligence".[14] It solidified the commitment to human-centric and 'trustworthy AI', adding the first step towards a future legislative framework built on the concept of 'trustworthy AI' to the new EU AI governance portfolio. The 'White Paper on AI' (European Commission, 2020d) is divided into two main sections, one on an Ecosystem of Trust, focussing on the first proposal for a regulatory framework, and one on an Ecosystem of Excellence, focussing on supporting the European AI ecosystem. Both of these closely match ambitions outlined in previous policy documents in Section I, such those in Europe's AI strategy (European Commission, 2018a), and fit in with other recent governance efforts. I will therefore use the 'White Paper on AI's' (European Commission, 2020d) duality of policy and infrastructure to highlight how far the EU AI policy has come in each area since this chapter's introductory Section and how they build on one another to make the EU a hub for 'trustworthy AI'.

On 21st April 2021, the European Commission published its package on a European approach for AI containing: a 'Communication on Fostering a European Approach to Artificial Intelligence' (European Commission, 2021a); a 'Coordinated Plan on AI: 2021 review' (European Commission, 2021b); and, the highly anticipated proposal for a 'Regulation on a European approach for Artificial Intelligence (AI Act)' (European Commission, 2021c). The 'Coordinated Plan on AI: 2021 review' (European Commission, 2021b) builds on the 'Coordinated Plan' (European Commission, 2018b) and dramatically expands its scope and ambitions, and the 'Regulation on a European approach for Artificial Intelligence (AI Act)' (European Commission, 2021c) builds on the 'White Paper on AI' (European Commission, 2020d) and subsequent impact assessments conducted by the

[14] See: https://ec.europa.eu/info/sites/default/files/political-guidelines-next-commission_en_0.pdf.

European Commission.[15] We will now see how all of this shapes up to form the context and the ecosystem for future AI governance in the EU.

## II.i. Trust and the EU's AI Governance

The chapter in the 'White Paper on AI' (European Commission, 2020d) dedicated to the Ecosystem of Trust outlined the European Commission's policy proposals for a potential regulation prior to the final publication of the proposal for a 'Regulation on a European approach for Artificial Intelligence (AI Act)' (European Commission, 2021c) on 21st April 2021. The Chapter was strongly inspired by the conceptual idea of 'trustworthy AI' and heavily referenced the work of the AI HLEG.

The core proposal suggested that in an envisioned horizontal legislation mandatory legal requirements should apply to high-risk cases of AI only. These high-risk AI cases were defined by the following cumulative criteria: if the sector itself is high risk (e.g. healthcare, transport and if the intended use involves high risk (e.g. injury, death, significant material/immaterial damage).

The mandatory legal requirements largely reflect the 'Ethics Guidelines' (AI HLEG, 2019) seven key requirements and are composed of the following: a requirement for adequate training data; a requirement for data and record keeping; a requirement for the provision of information; a requirement for robustness and accuracy; and, a requirement on human oversight. The final requirement was specifically laid out for the case of remote biometric identification. High-risk AI systems would be subject to conformity assessment (e.g. testing, inspection and certification) accounting for these requirements before they would be able to enter the EU market..

The 'White Paper on AI' (European Commission, 2020d) also outlined strategies for non-high-risk AI systems. It was suggested that these could partake in a voluntary labelling scheme which could build upon or implement the 'Assessment List' (AI HLEG, 2020). We see that the building blocks and vision sketched in Section I are starting to take considerable shape building the EU's AI governance future.

---

[15] See:
https://eur-lex.europa.eu/legal-content/EN/TXT/PDF/?uri=PI_COM:Ares(2020)3896535&from=EN.

Soon after, the legal affairs committee of the European Parliament adopted several aligned reports. These tackled an ethical framework for AI, civil liability claims against operators of AI systems and the protection of intellectual property rights with regards to AI.[16] It is noteworthy that the first report's guiding principles strongly resembled those of the 'Ethics Guidelines'. We can infer that the vision of EU AI governance is coherent across the EU institutions, which is important[17] as we move to the most recent and high-profile policy development on the European Commission's side: the 'Regulation on a European approach for Artificial Intelligence (AI Act)' (European Commission, 2021c).

Following the publication of the 'White Paper on AI' (European Commission, 2020d), the European Commission conducted impact assessments and opened a stakeholder consultation to receive feedback on the 'White Paper on AI' (European Commission, 2020d). This feedback shaped the subsequent proposal for a regulation.

The proposed 'Regulation on a European approach for Artificial Intelligence (AI Act)' (henceforth: 'AI Act'; European Commission, 2021c) introduces the European Union's legislation for AI, specifically high-risk AI systems. It is a risk-based regulation which covers stand alone AI systems that are considered high-risk which are elaborated on in Annex III to the 'AI Act' and cover use cases such as in law enforcement for individual risk assessment, education and vocational training for determining access to educational or training institutions or specific cases of access to essential public and private services and benefits. In short, Annex III lists a number of areas and specific use cases in those areas where stand-alone AI systems will automatically be considered high risk due to their potentially adverse impact on health, safety or fundamental rights of persons or groups. The other case of high-risk AI systems are those that are not stand-alone AI systems but those that are safety components of products or systems, or those that are products or systems.

[16] See: https://www.europarl.europa.eu/news/en/press-room/20200925IPR87932/making-artificial-intelligence-ethical-safe-and-innovative. The second legislative initiative is on 'liability for AI causing damage', focusing on civil liability claims against AI-systems, and the third report addresses intellectual property rights (IPRs) with relation to AI, suggesting that AI lacks a legal personality, and therefore inventorship should be exclusive to humans.

[17] The proposal for a Regulation on a European approach for Artificial Intelligence will need to pass through European Parliament and the Council. Once these two institutions agree on a final text, the regulation will be adopted.

Both of these types of high-risk AI systems need to comply with a number of requirements the 'AI Act' (European Commission, 2021c) lays down in Title III Chapter II, although the manner in which that compliance is achieved, documented and assessed (conformity assessment) is different between stand-alone and integrated high-risk AI systems. In the case of high-risk AI systems that are safety components of products or systems, or are themselves products or systems, the harmonised 'AI Act' (European Commission, 2021c) adjusts to fit with the existing sectoral procedures, rules and regulations.

The scope of this 'AI Act' (European Commission, 2021c) encompasses a range of actors: providers that place their AI system on the EU market, users of AI systems in the EU (except those that use it in a personal, non-professional activity) and providers and users of AI systems that are not based in the EU but where the output of their AI system is used in the EU.

All high-risk AI systems need to fulfil the requirements set out in the 'AI Act' (European Commission, 2021c) Title III Chapter II. These requirements closely match those that were previously proposed in the 'White Paper on AI' (European Commission, 2020d) and, as we have seen, are therefore closely connected to the requirements within the 'Ethics Guidelines' (AI HLEG, 2019). The requirements listed in the 'AI Act' are: *Data and Data Governance*; *Technical Documentation*; *Record Keeping*; *Transparency and Provision of Information to Users*; *Human Oversigh*t; and *Accuracy, Robustness and Cybersecurity*.

Whilst they are not described in this format, I would like to propose that the requirements can be thought of in two categories: those that are procedural and those that are informative. Data and Data Governance, Human Oversight and Accuracy, Robustness and Cybersecurity are procedural. They concern themselves with the workings of the algorithm throughout its lifecycle and how these can be affected in a positive manner to avoid negative impacts. By contrast, Technical Documentation, Record Keeping and Transparency and Provision of Information to Users can be considered as informational requirements.

They track procedural information, check it and monitor it throughout the AI system's life cycle.

Those actors that are responsible for ensuring that a high-risk AI system complies with the 'AI Act' have to fulfil certain conditions on top of adherence to the requirements mentioned previously. In short, they have to first build their AI system in accordance with the requirements from Title III Chapter II, then they have to undertake an internal conformity assessment of the AI system which encompasses paper trails and documentation generated in the first step and developed as a framework for the AI system throughout its functioning. That will entail a Quality Management System which ensures compliance with the 'AI Act' (European Commission, 2021c), a Risk Management system which acts as a continuous iterative process throughout the AI system's lifecycle, and Technical Documentation which covers elements such as detailed descriptions, pre-determined changes of the AI system and the performance, as well as monitoring, functioning and control of the AI system. Third, the provider or relevant other actor needs to establish a post-market monitoring system for the AI system once it has been put on the market or placed into service. This post-market monitoring system will collect logs produced by the AI system, act as a supervisor to the AI system and report serious incidents if they occur. Finally, before the AI system can be put on the market or placed into service it must be registered in the EU database, and EU Declaration of Conformity must be filled out to describe its adherence to the 'AI Act' (European Commission, 2021c) and it should be affixed with a CE marking to indicate that it has passed its conformity assessment.

In addition to requirements and procedures for high-risk AI systems, the 'AI Act' (European Commission, 2021c) also lays down a number of AI systems that are prohibited for use in the EU under certain conditions. Without enumerating them in detail, these prohibited AI system cover those that deploy subliminal techniques that could cause harm, those that exploit vulnerabilities in a manner that would cause harm and those that public authorities could use to evaluate the trustworthiness of an individual, leading to unfavourable treatments in different contexts or treatment that is disproportionate. Moreover, it includes 'real-time' biometric identification systems if they are used in publicly accessible spaces and for the purpose of law enforcement. However, noteworthy exceptions to the latter are cases where there is a targeted search for potential victims of crime, where it is in the public interest to prevent specific, substantial and imminent threats and to detect certain perpetrators.

Akin to the proposals in the 'White Paper on AI' (European Commission, 2020d), the 'AI Act' (European Commission, 2021c) also briefly concerns itself with voluntary Codes of Conduct for non-high risk AI systems, with the intention to foster 'trustworthy AI' and, therefore, compliance to the 'AI Act' within the broader ecosystem. The next Subsection will discuss how the corresponding environment within the EU is boosted in order to establish the overarching framework that these policy and legislative ambitions would fit in with. Section III will then concern itself in more detail with specific elements of the ecosystem that are likely to become crucial to the EU's success in AI governance in the near future.

## II.ii. Strengthening the AI Ecosystem

In recent years, it has become clear that the EU does not solely wish to rely on their regulatory expansionism, exporting norms and legislative approaches towards 'trustworthy AI' on an international stage. Acknowledging that its ecosystem has, at times, difficulty competing with tech giants developing or established outside of the EU, it is increasingly moving towards Digital Sovereignty. This encompasses the broader AI landscape in the EU. In order to have a truly comprehensive and integrated approach towards AI governance, ethical, policy and regulatory efforts must be boosted in tandem with the existing and foreseen landscape. In short, an increase in relevant EU infrastructure for AI development, deployment and use, equals an increase in ownership of the technology, an increase in the ability to shape it directly through norms for trustworthy AI (through soft and hard law) and a decrease in reliance on outside actors. Keeping this in mind, the following paragraphs will sketch how the EU is building this infrastructure and what benefit this may yield, starting with the chapter in the 'White Paper on AI' on an Ecosystem of Excellence and expanding it further with efforts outlined in the 'Coordinated Plan on AI: 2021 review' (henceforth: 'Coordinated Plan: 2021 review'; European Commission, 2021b) and adjacent initiatives.

Many of the areas below directly link back to aspects mentioned in Section I such in the 'Declaration' (European Commission, 2018c), the 'Coordinated Plan' (European Commission, 2018b) and in the AI Strategy (European Commission, 2018a), which all outlined the need to boost the ecosystem, to combine resources and to increase technical capabilities to ensure the EU's leadership in human-centric and 'trustworthy AI'.

The 'White Paper's' (European Commission, 2020d) chapter on an Ecosystem of Excellence concerns itself with technical infrastructure as well as with ecosystem building. On the latter, it especially focuses on building new infrastructures. On the more research-oriented side, it proposed work on establishing a lighthouse centre of research, innovation and expertise, to combat a seemingly fragmented AI research community in the EU. Looking at industry, small and medium-sized enterprises (SME) and start-ups, it calls for the development of testing and experimentation facilities (TEFs), building out capacity via the Digital Innovation Hubs (as mentioned in the 'Coordinated Plan'; European Commission, 2018b), a recently funded AI-on-Demand platform[18] and engagement of key stakeholders through a Public-Private Partnership on AI, data and robotics in the context of Horizon Europe.[19] It also touches on turning the 'Assessment List' (AI HLEG, 2020) into an indicative curriculum for those developing AI, and an ambition to keep talent and attract talent to the EU through new education networks under the Digital Europe programme.[20]

In accordance with the 'European Data Strategy' (European Commission, 2020a), the 'White Paper on AI' also puts an emphasis on the role of data in AI development ("compliance of data with the FAIR principles will contribute to build trust and ensure re-usability of data"; European Commission, 2020d), as well as the importance of computing infrastructure. This will be explored in more detail at the end of this section.

Every single one of these proposed efforts ties in with the 'Coordinated Plan: 2021 review' (European Commission, 2021b) and demonstrates the EU's efforts to build an infrastructure that can match its ambition on the governance side to promote the development of human-centric, sustainable, inclusive and trustworthy AI. While the

[18] See: https://cordis.europa.eu/project/id/825619.
[19] Horizon Europe is the current Multiannual Financial Frameworks programme and runs from 2021-2027 to support research, science and innovation with EUR 95.5 bn.
[20] See: https://digital-strategy.ec.europa.eu/en/activities/digital-programme.

‘Coordinated Plan’ (European Commission, 2018b) mapped out the initial areas in which the Member States should pool their resources and coordinate their actions, the ‘Coordinated Plan: 2021 review’ (European Commission, 2021b) moves towards an action-oriented approach with concrete joint actions focussing on the implementation of concrete measures and the removal of remaining fragmentation.

It is built around four key pillars: (1) to set enabling conditions for AI development and uptake; (2) to make the EU a place where excellence thrives from the lab to the market; (3) to ensure that AI works for people and society as a force for good; and (4) to build strategic leadership in high-impact sectors. These high-impact sectors  encompass areas such as smart mobility, law enforcement, migration and asylum, climate and the environment. In tandem with the third pillar, which encompasses a promotion of ‘trustworthy AI’ globally and nurturing of talent and skills, it could be seen as a reflection of the second component in the ‘Communication on AI in Europe, that is the EU’s initial AI strategy, which focussed on socio-economic changes associated with AI. Although the focus on scaling up the EU technical infrastructure is evidenced across all four pillars, the first is of particular relevance.

The policy document homes in on governance coordination frameworks and, crucially, on data infrastructures and computing capacities in order to create an enabling environment for AI development and uptake.

Referring back to the ‘European Strategy for Data’ (European Commission, 2020a), which aims to establish a single market for data within the EU and and the ‘Proposal for a regulation on European data governance (Data Governance Act)’ (European Commission, 2020e), which proposes several regulatory measures to increase society’s trust in data sharing, the ‘Coordinated Plan: 2021 review’ (European Commission, 2021b) outlines a number of core actions for data and an associated cloud infrastructure. These actions include establishing a new European Alliance for Industrial Data, Edge and Cloud,[21] co-investing with the Member States in common European data spaces and a European cloud federation and investigating the opportunity to set up an Important Project of Common European Interest (IPCEI) for next generation cloud infrastructures.

[21] See: https://digital-strategy.ec.europa.eu/en/library/cloud-and-edge-computing-different-way-using-it-brochure.

Prior to the 'Coordinated Plan: 2021 review' (European Commission, 2021b), at the end of 2020, 27 Member States signed a joint 'Declaration on Building the next generation cloud for businesses and the public sector' (European Commission, 2020c) where they expressed their intention to establish a secure, trustworthy and competitive cloud infrastructure in Europe for public administration, businesses and citizens alike. It addresses efforts aligned with those in the 'Coordinated Plan: 2021 review' (European Commission, 2021b), such as pooling of EU, national and private investment and shaping the process in accordance with the European Alliance on Industrial Data and Cloud,[22] and fostering technical solutions and policy norms for an interoperable pan-European cloud service.

Adjacent to this is Gaia-X,[23] an independent European platform for cloud infrastructure launched by France and Germany[24] intends to increase European cloud competitiveness vis-a-vis the US and China.

With an eye to infrastructure, the 'Coordinated Plan: 2021 review' (European Commission, 2021b) looks to support the development of High Performance Computing capabilities, as well as AI hardware. The latter encompasses investment in micro-electronics for AI chips, neuromorphic computing, photonics and projects under the Electronic Components and Systems for European leadership Joint Undertaking (ECSEL JU).[25] More specifically, actions call for the launch of an Industrial Alliance on Microelectronics, supporting research and innovation actions for low-power edge AI, and investing in processor and semiconductor technologies.

In fact, as part of its goal of achieving Digital Sovereignty, the EU quite evidently is aiming to advance its capabilities and to lessen its reliance on international actors when it comes to the design and production capabilities of low-power processors for AI and towards 2nm processor technologies. In late 2020, 18 Member States signed a 'Declaration on a European Initiative on Processors and Semiconductor Technologies' (European Commission, 2020b) to consolidate resources and boost the EU's electronics and embedded systems value chain.

---

[22] See: https://digital-strategy.ec.europa.eu/en/news/towards-next-generation-cloud-europe.
[23] See: https://www.data-infrastructure.eu/GAIAX/Navigation/EN/Home/home.html.
[24] See: https://www.euractiv.com/section/digital/news/digital-brief-the-gaia-x-generation/?utm_content=1591278775&utm_medium=eaDigitalEU&utm_source=twitter.
[25] See: https://www.ecsel.eu/what-we-do-and-how#:~:text=The%20ECSEL%20Joint%20Undertaking%20%2D%20the,era%20of%20the%20digital%20economy.

The 'Coordinated Plan: 2021 review' (European Commission, 2021b) also accounts for High-Performance Computing. It encourages Member States to continue developing large-scale High-Performance Computing infrastructure and references the importance of the EuroHPC Joint Undertaking.[26]

A proposed new independent Regulation for the EuroHPC[27] is expected to lead to an increase in the acquisition and development of supercomputers in the EU, rebalancing the scale in favour of the EU. Overall, it aims to develop exascale supercomputers with over '1 billion billion' operations per second (10^18 ops/second), to support the development of quantum and hybrid computers (also described in the 2018 regulation, section 21) and to create 33 'national competence centres' which will help to provide easier access to HPC opportunities locally and strengthen knowledge and expertise.

The second pillar in the 'Coordinated Plan: 2021 review's' (European Commission, 2021b) focuses on knowledge transfer and horizontal actions to support research and innovation (R&I). Its actions cover stakeholder collaboration, expanding and mobilising research capacities, building up suitable TEFs and funding AI solutions and ideas. Stakeholder collaboration will range from the Public-Private Partnership on AI, Data and Robotics[28] to a co-programmed European Partnership on Photonics,[29] supporting the EU's drive towards technological sovereignty. Whereas the European Commission already invested over €50 million in AI excellence centres through Horizon 2020,[30] it suggests funding more AI excellence centres and encourages Member States individually to set up regional and national excellence centres. Reminding ourselves of this chapter's earlier Section on the importance of 'trustworthy AI' for EU AI governance, it needs to be highlighted that the document suggests that funded programmes for AI under Horizon Europe are expected to adhere to the 'ethics by design' principle, including 'trustworthy AI'. We can see that various earlier threads are starting to come together. Finally, as mentioned in the 'White Paper on AI' (European Commission, 2020d) and the earlier 'Coordinated Plan' (European Commission, 2018b) Digital Innovation Hubs (DIHs) play a role in strengthening the ecosystem. To that end, alongside TEFs for specific sectors, such as edge AI or agri-food, the European Commission will support a scaling up of existing DIHs and set up new networks for what it terms European Digital Innovation Hubs (EDIHs) with AI expertise. These EDIHs will connect SMEs and start-ups with resources made available via the AI-on-Demand platform and relevant TEFs to make the AI system ready for deployment within the EU market. Finally, an investment of €1 billion from Horizon Europe and the Digital Europe programmes is expected between 2021-2027. The Digital Europe programme overall budget funds artificial intelligence (€2.1bn), HPC (€2.2bn) and cybersecurity (€1.7bn). The ambition remains the same as in the earlier 'Coordinated Plan' (European Commission, 2018b), namely to raise this to €20bn per year through public and private investment. Other institutions that are expected to fund AI in the EU are the European Innovation Council,[31] the European Investment Bank[32] (via the European Innovation Fund[33]) and the European Institute of Innovation and Technology.[34]

[26] See: https://eurohpc-ju.europa.eu/.
[27] See: https://ec.europa.eu/commission/presscorner/detail/en/ip_20_1592.
[28] See: https://ai-data-robotics-partnership.eu/.
[29] See:
https://www.photonics21.org/#:~:text=The%20European%20Technology%20Platform%20Photonics21,growth%20and%20jobs%20in%20Europe.
[30] See: https://digital-strategy.ec.europa.eu/en/news/towards-vibrant-european-network-ai-excellence.

One issue the EU has historically faced is that promising companies are often purchased by foreign companies before they reach their full potential, gobbling up talent, and knowledge in the process. Although this is not an explicit part of strengthening the EU's AI landscape nor mentioned in the 'Coordinated Plan: 2021 review' (European Commission, 2021b) it is relevant to quickly mention the EU's regulatory framework to screen foreign direct investment (FDI).[35] This FDI framework aims to protect the EU's strategic interests and came into force at the end of 2020. Of particular relevance, in light of the EU's shift towards achieving digital sovereignty[36] and scaling its technical infrastructure, is that this framework covers assets that are 'critical technologies and dual use items'.[37] This includes amongst others AI, robotics and semiconductors.

---

[31] See: https://eic.ec.europa.eu/index_en.
[32] See: https://www.eib.org/en/index.htm.
[33] See: https://ec.europa.eu/clima/policies/innovation-fund_en.
[34] See: https://eit.europa.eu/.
[35] See: https://eur-lex.europa.eu/legal-content/EN/TXT/?uri=celex%3A32019R0452.

From Sections II.i and II.ii it is evident that the EU is both (1) serious in its pursuit to strengthen its vision of human-centric 'trustworthy AI' by shaping the AI governance framework through regulation, policy and certification; and, (2) willing to build out the entire ecosystem in support of this vision, positioning itself as a future sovereign digital actor and third way between the US and China.

# III. The Future

## Sketching the future of AI governance in the EU

In the previous two Sections we saw how the EU built up its strategy and how all of the elements fit in with the larger tapestry of the EU's approach to AI governance. Section III will be based on the current dynamic that the EU is exhibiting and briefly sketch three AI governance areas that are prime candidates to become crucial for AI governance in the EU in the coming decade.

The finalisation and implementation of the 'AI Act' (European Commission, 2021b) over the coming months and years is a clear candidate, but there are other less obvious but equally relevant ones. The following paragraphs pick them out, polish them and highlight their importance in the future pathways for EU AI governance.[38]

---

[36] See: https://ec.europa.eu/info/strategy/priorities-2019-2024/europe-fit-digital-age_en.
[37] As defined in Article 2.1 of Regulation (EC) No 428/2009.
[38] It should be noted that this section is the personal opinion of the author and the likelihood of the proposed sketches coming into fruition varies.

# AI Megaprojects: a CERN for AI and AI lighthouses

Multiple experts have called for megaprojects within the EU over the past few years. Most notably, for a CERN for AI.[39] Certainly, such a project would be very ambitious. Nevertheless, upon careful reading of recent EU policy documents and the general drive to boost the technical landscape and ecosystem (as outlined in Section II.ii) a budding AI megaproject may be on the cards.

There are roughly two shapes such a project could take: being centralised within a Member State who has suitable technical infrastructure and is well located or broadly distributed across Member States, with one centralised headquarter. Given the existing ecosystem, ongoing efforts to boost research capacity and technical infrastructure and recent advocacy from large research groups within the EU,[40] both options may be viable.

As described earlier, the EU plans to build out their existing DIHs into EDIHs with significant focus on AI EDIHs. This would lead to a stark increase in the number of facilities where research can be conducted and where AI systems can be developed and experimented upon by SMEs and start-ups. Moreover, the European Commission has recently funded a large scale project called ELISE,[41] the European Network of AI Excellence Centres. ELISE collaborates with the European Laboratory for Learning and Intelligent Systems (ELLIS), a large network of European researchers, and closes the gaps between AI institutes in Europe. This adds to a previously funded network, TAILOR[42] (Foundations of Trustworthy AI - Integrating Learning, Optimization and Reasoning) whose goal is to create a network across Europe on the "Foundations of Trustworthy AI".

---

[39] See: https://www.steven-hill.com/why-we-need-a-cern-for-ai/.

[40] See: https://claire-ai.org/wp-content/uploads/2020/02/CLAIRE-Press-Release-11.pdf; https://www.timeshighereducation.com/news/scientists-split-europe-paves-way-cern-of-ai; https://sciencebusiness.net/news/call-cern-ai-parliament-hears-warnings-risk-killing-sector-over-regulation.

[41] See: https://cordis.europa.eu/project/id/951847.

[42] See: https://liu.se/en/research/tailor/about.

Efforts such as these evidence that there is fertile ground to establish a centralised large-scale headquarter from an increasingly powerful network and quasi-independent nodes.

Another key reason for why a large-scale AI project might be both on the cards and meaningful for the EU to establish can be found in the 'AI Act' (European Commission, 2021b). In tandem with new regulatory requirements, more TEFs will be needed. This ranges from TEFs specialised to test and assess for specific aspects of the conformity assessment, as well as those that can assess an AI system's entire regulatory fitness. Building out existing facilities for testing and experimentation and establishing novel ones will eventually lead to a dense landscape of distinct but similar institutions across the EU's Member States. It might be in the EU's best interest to centralise these facilities and locate them alongside big industrial efforts such as the European Cloud efforts, European Data Spaces and the HPC Joint Undertaking. Such a localisation could: increase efficiency and provide economies of scale for using data, research engineering, and other supporting infrastructure; enable more ambitious research, testing and experimentation efforts; and, encourage a lasersharp alignment between policy and practice.

Indeed, the European Commission has indicated ambitions to develop something akin to a CERN for AI in several policy documents, for example in the 'White Paper on AI' (European Commission, 2020d), and most recently in the 'Coordinated Plan on AI: 2021 review' (European Commission, 2021b). In particular, the development of AI Lighthouse Centres (or, a centre) within the EU is championed. This would be a large-scale research facility for AI.

It would be promising for the EU's ambitions on a global playing field to establish an AI lighthouse centre, a CERN for AI or another version of a large-scale facility for AI research and development. Most importantly, this could lead to the EU becoming a truly unified player where fragmentation between various European research institutes is superseded (Stix, 2018), significant chunks of the aimed for EUR 20bn per year funding for AI could be centralised for ambitious projects, and new talent could be attracted to the EU (Stix, 2019).

Considering a future landscape with an increasing number of networked institutions, mounting calls from large AI research networks within the EU, and the policy proposals from the European Commission, the future of EU AI governance may well hold an AI megaproject.

## AI Agencies: regulation, measurement and foresight

The idea of a large European AI Agency is not new. The very first document presented in this chapter, the resolution on 'Civil Law Rules on Robotics' (European Parliament, 2017), already called for the establishment of an EU Agency for Robotics and AI in "order to provide the technical, ethical and regulatory expertise needed to support the relevant public actors, at both Union and Member State level, in their efforts to ensure a timely, ethical and well-informed response to the new opportunities and challenges, in particular those of a cross-border nature".

Similarly, the 2019 European Parliament report 'A comprehensive European industrial policy on artificial intelligence and robotics' (European Parliament, 2019) called for the establishment of a European regulatory agency for AI and algorithmic decision-making. With this backdrop, and tracing the institutional landscape mapped out by the 'AI Act' (European Commission, 2021c) it is likely that the EU will eventually establish a new institution, specifically for the governance of AI. The 'AI Act' (European Commission, 2021c) envisions a complex institutional interplay to sustain the regulatory measures for AI. This encompasses various national institutions: those that fall under the National Competent Authorities, which would be the National Supervisory Authority, the Notifying Authority and various Notified Bodies (official conformity assessment bodies); Market Surveillance Authorities; and, from the European Commission's side a novel European AI Board (where e.g. member states will be represented) and an expert group. All of these will play a crucial role for the application of the horizontal regulation for AI within the EU and will have different scopes and powers. Some will have investigative power and some will assess the suitability of AI systems for the European market. Of course, many of these institutions cannot (and should not) be merged. Nevertheless, after an initial phase of getting to know the ropes of the final agreed upon regulation, it is likely that there will be a time window in which an EU AI Agency would be built to combat fragmentation, pool expertise and streamline various workflows.

Beyond the aforementioned scopes, the ‘AI Act’ (European Commission, 2021c) also has a provision which ensures that new AI systems can be added to the list of high-risk AI systems as and when deemed appropriate. In order to ensure that timeliness and foresight are underlining this power, and more generally, to ensure that policy making matches technological progress, I propose that another version of an EU AI Agency -- which has not been part of any EU-level discussions yet -- should be considered: a European AI observatory.

Historically, observatories were established to measure and survey natural occurrences e.g. astronomical, geophysical or meteorological events. An AI observatory as envisaged here, on the other hand, would monitor, measure and benchmark AI progress, a technology created by humans.

Although the EU is involved in OECD efforts towards an international AI policy observatory and has its own body, the AI Watch,[43] this does not yet live up to what an EU AI observatory could look like. As envisaged here, a EU AI observatory should have the capacity and ambition of conducting independent forecasting and measurement exercises. These in turn would ensure that policy making, regulation and other governance efforts in the EU are aligned with the technical state of the art of AI systems and sufficiently future proof.

As previously indicated, in order for policy makers and regulators to make suitable and timely decisions to add potential future high-risk AI systems to the ‘AI Act’ (European Commission, 2021c), either to regulate them or to ban them, they need to be aware of ongoing technical developments. One way of doing so from a government’s perspective could be to monitor the technical landscape, measure technical progress, and therewith notice crucial shifts that could indicate a cause for concern or intervention.

[43] A joint initiative between the European Commission’s Joint Research Centre (JRC) and the Directorate General for Communications Networks, Content and Technology (DG CONNECT). See more: https://knowledge4policy.ec.europa.eu/ai-watch/about_en.

Overall, AI governance in the EU, through regulation, standards, certification or other efforts could be significantly more impactful, agile and anticipatory if it narrowed the pacing gap between technological progress and governance efforts (Marchant et al., 2011). Furthermore, metrics can be seen as comparatively non-threatening and could encourage information sharing between countries and institutions, indirectly promoting collaboration and cooperation.

Taking these aspects into consideration, I suggest that an EU AI observatory would be vastly beneficial to the EU's AI ambitions and could be seen as a contributing factor for better regulatory measures in the future.

## Standards

Lastly, standards will play an important role in the future of the EU's AI governance. While standardisation efforts are not a 'European-only' effort, they are likely to meaningfully shape the EU market for AI systems. Crucially, the 'AI Act' (European Commission, 2021c) notes that if suitable standards exist[44] that would cover one, or more of the relevant legal requirements for an AI system to pass conformity assessment  (outlined in Section II.ii.) then an adherence to those standards can be considered as an adherence to the legal requirement(s) in question. Of course, if a provider chooses not to follow an existing standard or, where a standard does not exist then they must prove suitable and sufficient adherence to the legal obligations in a different manner. This goes to say that actors involved in standardisation bodies and those directly working on standards will have some non-negligible leeway in shaping the future mechanisms with which many of those adhering to the regulatory framework for AI will tackle their conformity assessments. Standardisation efforts might framing some of the hurdles a high-risk AI system needs to pass before it enters the EU market and shape the manner in which relevant actors think about assessing high-risk AI systems.[45] I propose that it is a key lever for upcoming governance measures within the EU.

[44] Those standards would have to be published in the Official Journal of the European Union.

# IV. Conclusion

In conclusion, this chapter first introduced the background of the EU's AI governance ambitions, drawing together the different elements and highlighting how they interconnect, together developing the tapestry out of which the EU's vision and current governance efforts result. Subsequently, it introduced the two-pronged recipe the EU pursues for AI. There, it discussed the corresponding roles of 'trustworthy AI' and regulatory efforts with the associated scaling of the technical infrastructure within the EU. Together, it demonstrated how these choices support the EU's ambition to be a leader on ethical and human-centric AI on an international stage. Finally, the author sketched three possible future directions they envision the EU moving towards: AI research megaprojects, new AI Agencies, and an increasing importance of standardisation efforts.

[45] Assuming the provider chooses to use standards or technical specifications for their conformity assessment. However, if standards are available it is likely that most providers will choose to adhere to the standards to streamline and minimize their workflows between various geographical regions.

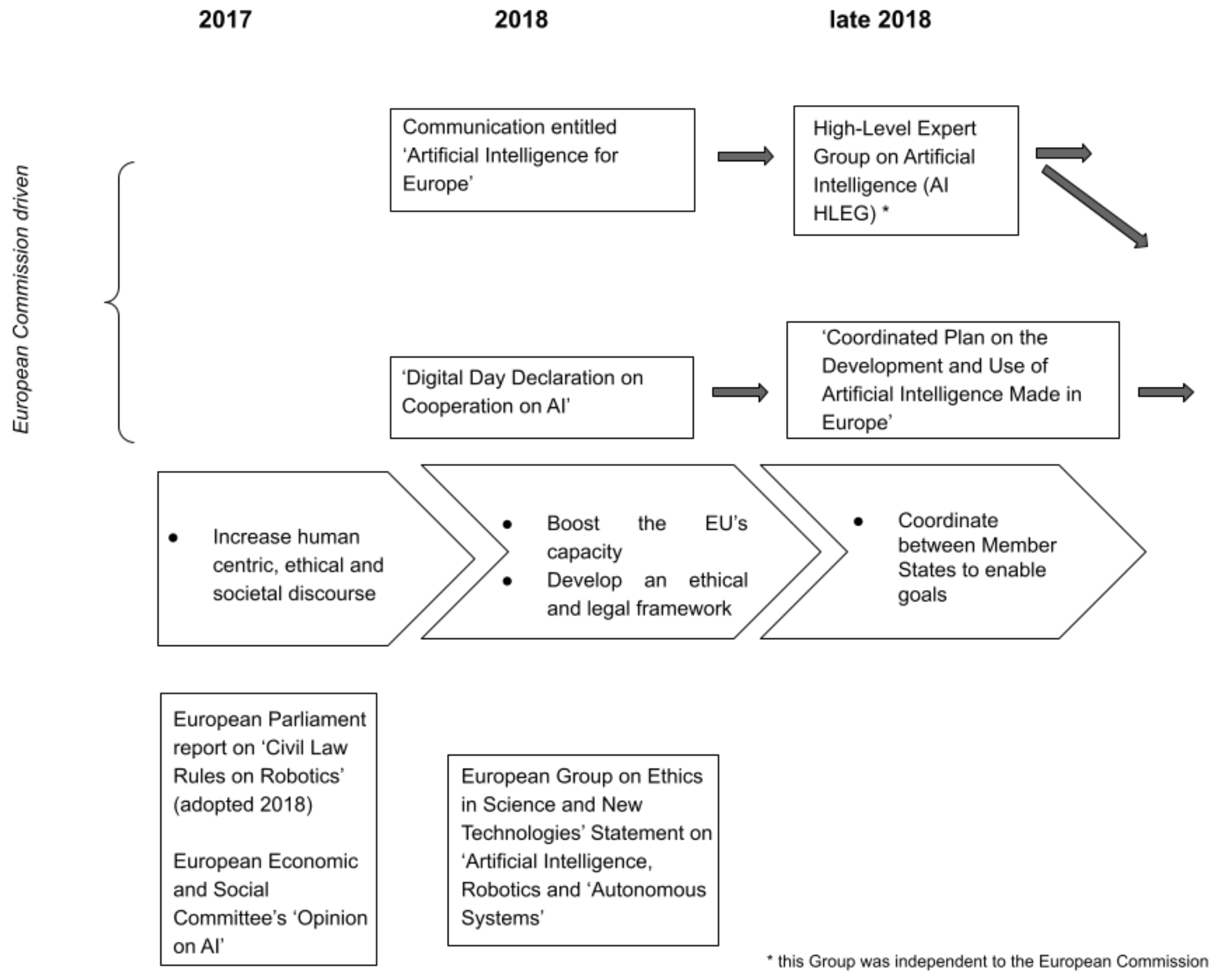

*Figure 1a: Flowchart of several of the documents outlined in this chapter and some of their interconnections.*

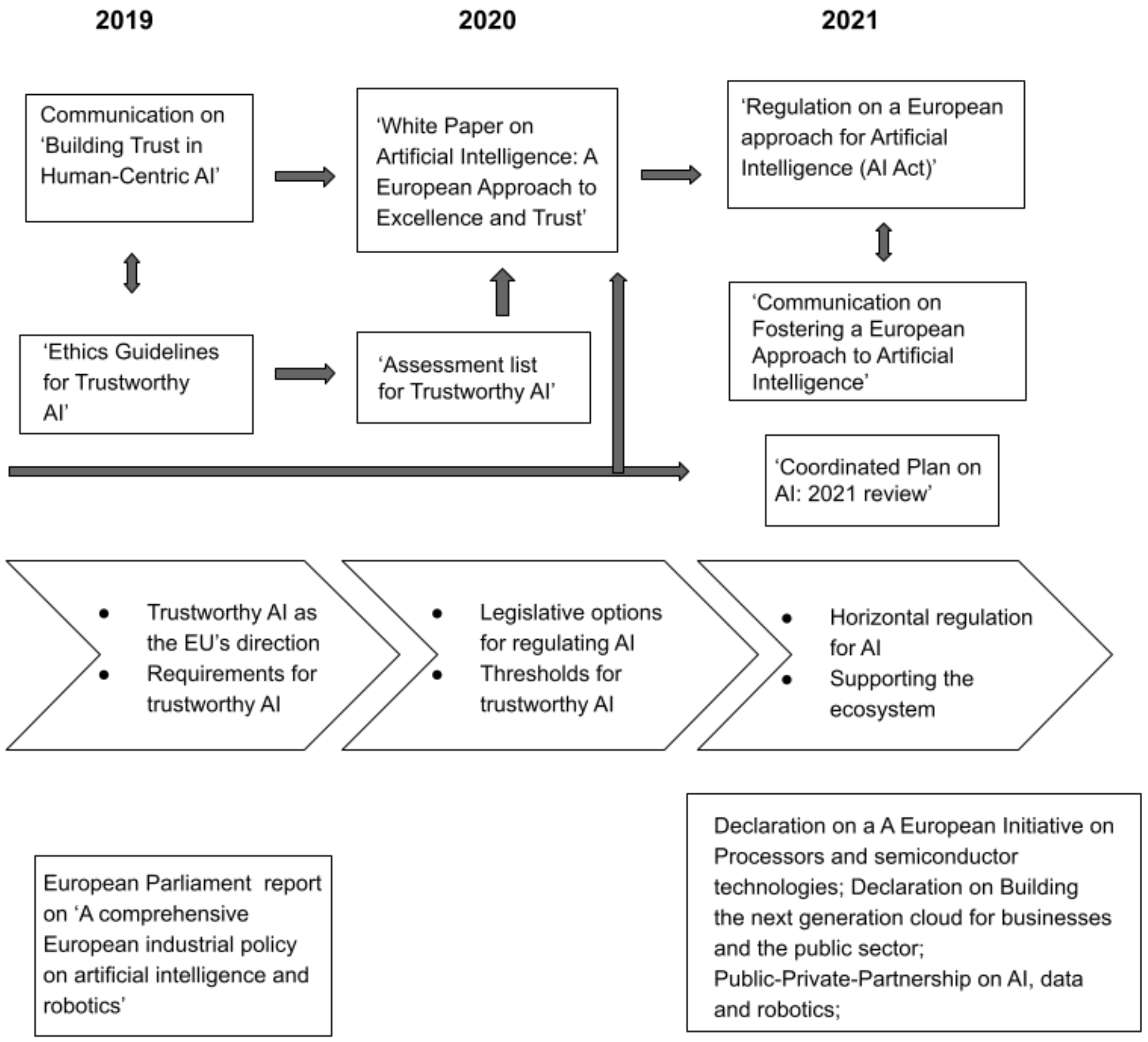

*Figure 1b: Flowchart of several of the documents outlined in this chapter and some of their interconnections.*